\documentclass[letter]{aa} 

\usepackage{natbib}
\bibpunct{(}{)}{;}{a}{}{,}
\usepackage{graphicx}
\usepackage{txfonts}
\usepackage[]{hyperref}
\usepackage{tikz}
\usepackage[]{xcolor,soul}

\usepackage[normalem]{ulem}

\usepackage{amsmath,amssymb,latexsym}

\usepackage[]{xcolor,soul}

\definecolor{Celestial}{HTML}{3C9AC9}

\usepackage{comment}

\begin{document}

\title{Extremely iron-poor O-type stars in the Magellanic Bridge}

\author{E.C. Sch\"osser\inst{1} 
\and V. Ramachandran \inst{1} 
\and A.A.C. Sander \inst{1} 
\and J.S. Gallagher \inst{2,3} 
\and M. Bernini-Peron \inst{1} 
\and G. González-Torà \inst{1} 
\and J. Josiek \inst{1} 
\and R.R. Lefever \inst{1} 
\and W.-R. Hamann\inst{4} 
\and L.M. Oskinova\inst{4}
}      

\institute{Zentrum f\"ur Astronomie der Universit\"at Heidelberg, Astronomisches
  Rechen-Institut, M\"onchhofstr. 12-14, 69120 Heidelberg, Germany
  \\ \email{elisa.schoesser@uni-heidelberg.de} \and Department of Astronomy, University of Wisconsin, 475 N. Charter St., Madison, WI 53706, USA \and
  Department of Physics and Astronomy, Macalester College, 1600 Grand Ave, St. Paul, MN 55105 USA \and
 Institut f{\"u}r Physik und Astronomie, Universit{\"a}t Potsdam, Karl-Liebknecht-Str. 24/25, 14476 Potsdam, Germany
}
\date{}

\abstract {To study stars analogous to those in the early Universe with redshift $z>3$, we need to probe environments with low metallicities. Until recently, massive O-type stars with metallicities lower than that of the Small Magellanic Cloud (SMC, $Z<20\% Z_\odot$) were only known in compact dwarf galaxies. Observations of stars in such distant galaxies ($> 1$ Mpc) suffer from limited signal-to-noise ratios and spatial resolution. Recently, a few O-type stars were identified in the nearby Magellanic Bridge which offers a unique laboratory with low gas density and metal content.}
{We acquired high-resolution HST-COS FUV spectra of two O-type stars in the Magellanic Bridge. Using the UV  forest of iron lines from these observations, we aim to precisely measure the inherent iron abundances and determine the metallicity of the stars.
}
{Using detailed expanding non-LTE atmosphere models, we generate synthetic spectra for different iron abundances and for a range of microturbulent velocities. We use Bayesian posterior sampling to measure the iron abundance and compute uncertainties based on the possible range of microturbulent velocities.}
{The O stars in the Magellanic Bridge have severely sub-SMC iron abundances, reaching as low as 10.8\% and 3.6\% $\mathrm{Fe}_\odot$. The most Fe-deficient star also shows $\alpha$-enhancement. These stars are the nearest extremely metal-poor O stars discovered to date.}
{Our finding marks the first robust determination of O-star iron abundances in a metallicity regime comparable to dwarf galaxies like Sextans A and Leo P. The iron abundances of the stars do not correlate with their oxygen abundances. Our results highlight the problem of using oxygen-based metallicities. The proximity of the stars in the Bridge combined with their different abundance patterns underlines that the ISM of the Magellanic Bridge must be highly inhomogeneous and is not properly mixed.}


\maketitle

\section{Introduction}

Massive stars ($M_{\mathrm{init}} > 8 \, M_\odot$) play an important role for the chemical enrichment and ionization of the Universe. 
Following the Big Bang, the first Population III stars formed from pristine gas clouds which almost entirely consisted of hydrogen and helium \citep[e.g.][]{2002Bromm}.
Through nuclear fusion, these hot massive stars produced heavier elements (metals), primarily the $\alpha$-peak  elements, which were dispersed into the interstellar medium (ISM) via intrinsic mass loss
and their final core-collapse supernovae \citep[see e.g.][for a review]{2002Woosley,2007Janka}, seeding future generations of star formation.

Studying metal-poor massive stars, born in environments with minimal prior chemical enrichment, provides a unique window into the early stages of stellar evolution and nucleosynthesis.
These stars are local analogs of the metal-poor stellar populations observed in high-redshift galaxies by the James Webb Space Telescope (JWST).
The increasingly rich JWST spectroscopic observations of metal-deficient galaxies at high redshifts highlight the lack of a deeper quantitative understanding and problems in current models at low metallicity \citep[see, e.g.,][]{2024Garg}, e.g., when interpreting observed emission line ratios from ionized gas \citep{2023Matthee,2024Laseter,2025Rowland}.
As discussed by \citet{2023Hopkins} the interpretation of spatially integrated JWST spectra of most sources are expected to be affected by a variety of issues, such as dust absorption and ISM structure, as well as properties of the stars. Massive star models grounded by observational data are essential for the analysis of optical spectra of young, low-metallicity galaxies at high redshifts.

Empirical constraints from individual, observed metal-poor massive stars predominantly stem from spectroscopic studies of stars in the Small Magellanic Cloud (SMC) with a mean metallicity of $Z \sim 20\% \, Z_\odot$ \citep[e.g.][]{2003Bouret, 2019Ramachandran, Rickard2022} 
or within the recent ``XShootU'' collaboration \citep[][]{2023Vink,2024Backs,2024BerniniPeron}.
However, JWST observations of galaxies at redshift $z>3$ reveal even lower metallicities up to $\sim$$2\% \, Z_\odot$ \citep[e.g.,][]{2024Curti,2024Boyett}.
Therefore, it is essential to spectroscopically analyze stars in environments with sub-SMC metallicities to bridge the gap between theoretical predictions and high-$z$ observations.
While old, low-mass, and extremely metal-poor ($Z < 10\% \, Z_\odot$) stars can be found in the Milky Way and Galactic halo \citep[e.g.][]{1988Gratton,2008Cohen}, young massive stars are missing there. A few massive stars with comparably low metallicities have been studied in comparatively distant ({$\gtrsim 1$~Mpc}) compact dwarf galaxies such as Sextans A and Leo P \citep[e.g.,][]{2019Evans,2019Garcia,2024Telford}.  
The observations of individual massive stars in such distant locations suffer from limited signal-to-noise ratios and spatial resolution.

The Magellanic Bridge, first discovered by \citet{1963Hindman}, is a stream of gas and mostly young stars connecting the two Magellanic Clouds. 
With its nearby distance of $\sim$$55$\,kpc \citep{2016Jacyszyn}, low foreground extinction, low density, and low mean metal content \citep{2001Lehner}, the Magellanic Bridge is a unique alternative laboratory in our cosmic neighborhood to study star formation and feedback at extremely low metallicity. Notably, \citet{2008Dufton} reported B stars in the Bridge with very low iron abundances.

The Magellanic Bridge contains a population of young stars (primarily B and A types) \citep{1999Rolleston}, but only recently the first three O-type stars were discovered \citep{Ramachandran2021}.
Two of them were found to have extremely low mean CNO abundances of 4\% and 8\% solar.
However, a main element affecting the mass loss and therefore evolution of massive stars is iron \citep[e.g.,][]{Abbott1982,ChiosiMaeder1986}. The iron abundance and opacity set the necessary conditions to launch the wind of massive stars and thus are the main ingredient of the scaling of the mass-loss rate with metallicity \citep[e.g.,][]{Pauldrach1987,2001Vink}.
However, in contrast to most metals that are promptly expelled to the interstellar medium during the core collapse of massive stars, iron is mainly produced with a considerable time delay during supernovae type Ia. 
Thus, the iron abundance does not necessarily scale with other metal abundances. 
Differences between $\alpha$/Fe have also been discovered for some massive stars, e.g., in IC\,1613 \citep{2014Garcia}.
For hot OB stars, determining the iron abundance requires high-resolution UV spectra and is often unfeasible for distances larger than the Magellanic Clouds. Rapid rotation, common at low metallicity further complicates analysis by smearing iron lines. Thus, determined iron abundances are often limited to cool supergiants with iron lines present in optical spectra \citep{2023Urbaneja}.

In this work, we use newly acquired HST UV spectra to measure for the first time the inherent iron abundances for the two O-type stars MBO2 and MBO3 in the Magellanic Bridge \citep{Ramachandran2021}, which show low mean CNO abundances and slow rotation. For this, we employ atmosphere models with low iron abundances and a newly developed Bayesian analysis framework.

\section{Observations}
The UV spectra (ID. \#16647, PI Ramachandran) of the two CNO metal poor O-type stars were taken with the grating G160M/1623 by the Cosmic Origins Spectrograph (COS) \citep{Green2012} onboard the Hubble Space Telescope (HST).

To assess the consistency of our approach, we re-examined the iron abundance of one of the sample stars from \citet{2008Dufton} (DI\,1388) which is reported to have the lowest iron content of only $X_{\mathrm{Fe}}=1.26\%\,X_{\mathrm{Fe,\odot}}$. 
Instead of the HST/GHRS/G200M spectra ($1888-1929\,\AA$) used by \citet{2008Dufton}, our analysis is based on STIS/E140M spectra (ID. \#7511, PI Keenan), which have a higher resolution and signal-to-noise ratio.

\begin{figure}
    \centering
    \includegraphics[trim=0 0 0 50, clip,width=\linewidth]{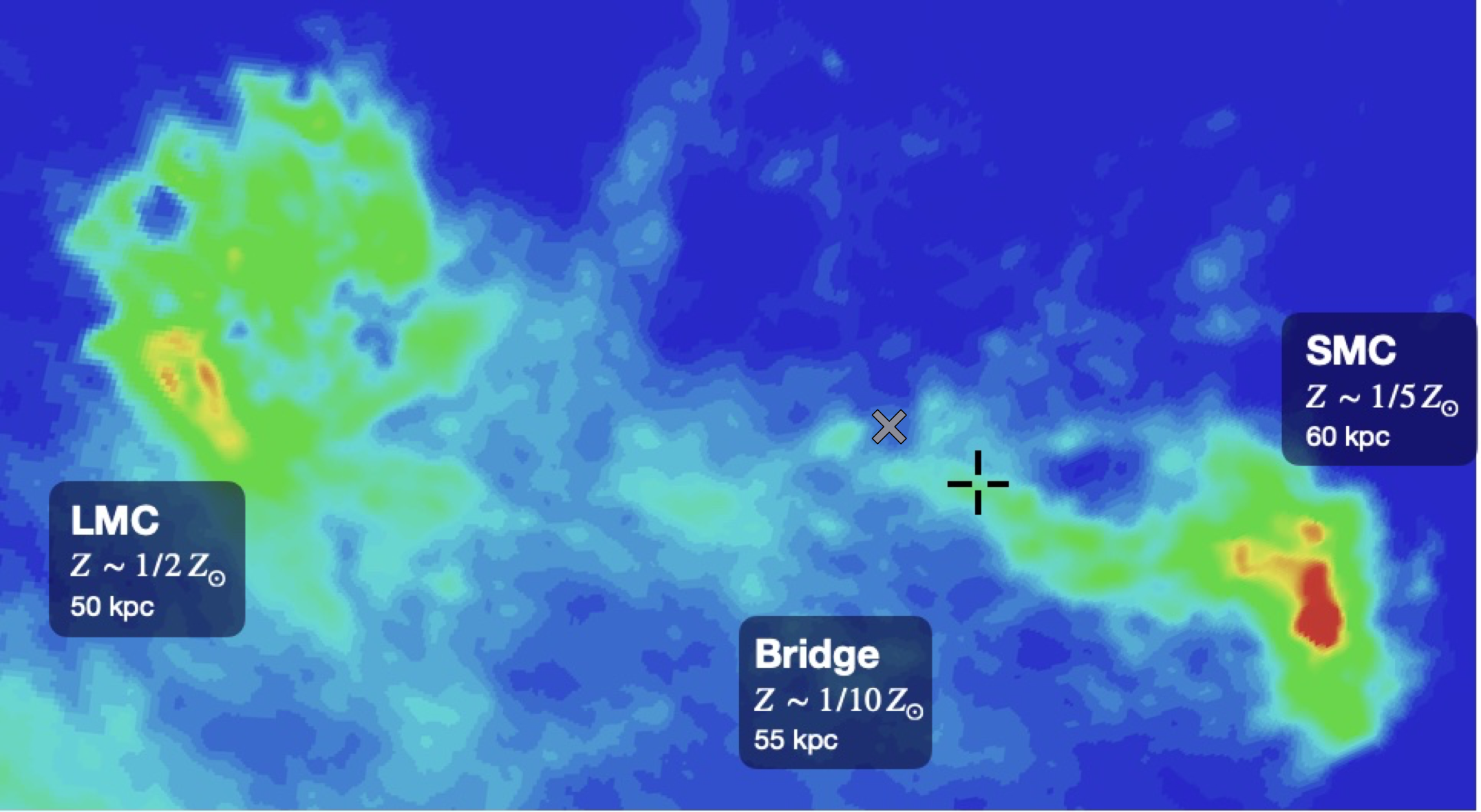}
    \caption{Neutral hydrogen map of the Magellanic Clouds and the Magellanic Bridge from the HI4PI survey \citep{2016HI4PI}. The black plus symbol marks the approximate position of MBO2 and MBO3 within the Magellanic Bridge, and the grey cross for DI1388.}
    \label{fig:mbrige-location}
\end{figure}

A summary of the stars' coordinates, spectral types \citep[from][]{Ramachandran2021}, and observational settings can be found in Table \ref{tab:coordinates}. 
In Figure \ref{fig:mbrige-location}, the location of the analyzed stars within the Magellanic Bridge is highlighted in respect to the SMC and LMC. 
The basic data reduction (e.g.,  normalization, radial velocity correction) is described in Appendix\,\ref{sec:app-basicred}.

\section{Data Analysis}

\subsection{Stellar Atmosphere Models}

For the spectral analysis, we use the PoWR non-LTE expanding stellar atmosphere code \citep[see, e.g.,][]{Grafener2002,2003Hamann,2015Sander} which calculates model atmospheres and generates synthetic spectra for a chosen set of wind and stellar parameters. 
PoWR accounts for line blanketing by grouping thousands of atomic energy levels from iron group elements into superlevels \citep{Grafener2002}.
Adopting the parameters (such as $T_\text{eff}$, $\log g$, $\varv \sin i$,  $\varv_{\mathrm{mac}}$ etc.) and abundances from \citet{Ramachandran2021} constrained from the optical regime, we calculate models with different iron abundances between 1 and 15~\% and varying microturbulent velocities (see Table \ref{tab:datom} for details on the used atomic data).
Instead of the iron mass fraction directly, PoWR uses the iron group abundance (herein denoted as $G$) as input parameter, consisting of scandium, titanium, vanadium, chromium, manganese, iron, cobalt, and nickel.
The specific elemental abundances are calculated internally, weighted by their relative solar abundances.
We do not analyze the wind features in this work, but instead employ fixed values for the terminal wind velocities of $\varv_\infty = 1994\,\mathrm{km\,s^{-1}}$, and mass-loss rates of $\dot{M}=10^{-9.0}\,M_\odot\,\mathrm{yr}^{-1}$ and $\dot{M}=10^{-8.5} \,M_\odot\,\mathrm{yr}^{-1}$ for MBO2 and MBO3, respectively. 
Throughout this work, our reference for the solar abundances are the values from \citet{2009Asplund} and \citet{2015Scott,2015Scott2}.

\subsection{Determination of the microturbulence}

Both stars, MBO2 and MBO3, have very narrow lines.
For MBO2, \citet{Ramachandran2021} could only determine upper limits for the microturbulence $\xi$, projected rotation velocity $\varv\,\sin i$ and the macroturbulence $\varv_{\mathrm{mac}}$ due to the extremely narrow absorption lines.
As there is a known degeneracy between Fe abundance and $\xi$ \citep[e.g.,][]{2015Bouret} when using the UV iron forest to determine Fe abundances, we carefully examined a range of microturbulent velocities between $1$ and $20\,\mathrm{km\,s^{-1}}$ that still reproduce the shape of the observed metal lines. 

\begin{figure}[htp]
    \centering
    \includegraphics[width=0.95\linewidth]{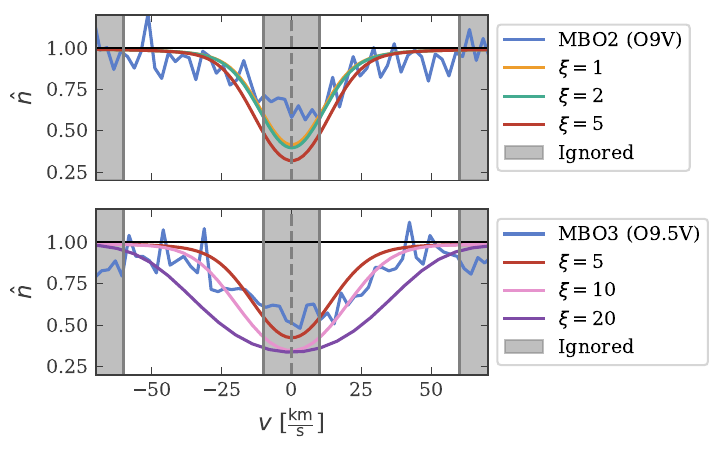}
    \caption{Comparison of the observed and modeled \ion{S}{v}$\lambda1501.76$ line for different $\xi$ for MBO2 (top) and MBO3 (bottom) in velocity space. The defined wing region in which we evaluate $\chi^2$ is highlighted.}
    \label{fig:vmic_comp}
\end{figure}

To find out which range of microturbulences agree with the observed spectrum, we compare the observed \ion{S}{v}~$\lambda$1501.8 line with models calculated varying $\xi$ (see Fig.\,\ref{fig:vmic_comp}). 
We use the \ion{S}{v} line as it is the strongest and most isolated metal line in the considered UV spectral range.
Microturbulence influences the broadened wings of these metal lines. 
While for MBO2, all models with $\xi \leq 5~\mathrm{km\,s^{-1}}$ match with the wings of the absorption lines, viable microturbulence values for MBO3 are in the range of $\xi=5-10~\mathrm{km\,s^{-1}}$. 
\citet{Ramachandran2021} determined a microturbulence of $\xi\leq2~\mathrm{km\,s^{-1}}$ for MBO2 and $\xi=8~\mathrm{km\,s^{-1}}$ for MBO3, thus agreeing with these results.
To later combine the fitting results for different $\xi$, we also quantify the fit to the \ion{S}{V} line by computing the $\chi^2$ value (see Appendix \ref{app:microturb}).

\subsection{Bayesian inference of iron abundances}\label{sec:bayes}
Due to the difficulty of visually distinguishing Fe lines from the noise level, a more advanced approach than fitting by eye is required for accurate Fe abundance determination.
As described in Appendix\,\ref{sec:app-basicred}, normalizing the UV spectra with synthetic models can require a small flux shift to the model continuum, corresponding to a slight change in stellar luminosity.
Using Bayesian posterior sampling, we infer both the flux shift and the Fe group abundance $G$ by fitting the forest of Fe\,\textsc{iv-v} lines for a range of fixed $\xi$ values. 
To sample the posterior distribution, we employ the Markov Chain Monte Carlo (MCMC) algorithm \texttt{emcee} \citep{2013Foreman}, assuming a Gaussian likelihood 
\begin{align}
     \mathcal{L} = \prod_{i=1}^{N} \frac{1}{\sqrt[]{2\pi}\sigma_{\mathrm{i}}} \, \mathrm{exp}\left( - \frac{(F_{\mathrm{obs, i}}- F_{\mathrm{sim, i}})^2}{2\sigma_{\mathrm{i}}^2}\right).
\end{align}
and using the estimated flux errors $\sigma_i$ from the COS data products \citep{2021Johnson}.

As the normalization varies across wavelengths, likely due to instrumental effects, and fitting the entire spectrum would require multiple shifts, we carefully restrict our fits to regions with reliable normalization.
Additionally, we ignore the parts of the spectrum where the noise dominates, and spectral regions with wind lines or other strong absorption lines. 
These ignored wavelength intervals are different for both stars (see Appendix \ref{app:detailed_results}) but are kept constant for all fits with different $\xi$.

We calculate a grid of PoWR models for each star with different iron metallicities, 8 models spaced between $1 \%$ and $15 \% \, G_\odot$ for MBO2 and 5 models between $1 \%$ and $9 \% \, G_\odot$ for MBO3, and otherwise fixed stellar parameters to the values determined from the optical spectrum in \citet{Ramachandran2021}.
To evaluate the spectrum at all metallicities within the grid, we use linear interpolation between the points.
We use uninformative prior distributions: For the iron metallicity, we use a uniform prior in the interval given by the grid, and for the flux shift we use a truncated Gaussian prior centered around $\mu_{\mathrm{shift}}=0$ and with standard deviation $\sigma_{\mathrm{shift}}=10^{-14}$
in the interval $\left[-3\times10^{-14},3\times10^{-14}\right]~\mathrm{erg\,cm^{-2}\,s^{-1}\,\AA^{-1}}$.
For each star, we run 10 chains with 10,000 steps.
To confirm that the fits have converged, we check that the chain length is sufficiently ($>$50 times) larger than the autocorrelation length $\tau$.
We discard the first 1,000 burn-in steps and thin the samples so that we only consider each $\tau/2$ step.

Finally, we compute the Fe abundance from the fraction of Fe in the G element (93\%) and the total metallicity $Z$ using the determined abundances for C, N, O, Mg, and Si from \citet{Ramachandran2021} plus our determined Fe abundance. 
The only missing element with a sufficient impact on $Z$ is neon.
As there is no Ne line detectable in the observed spectral range of the bridge stars, we estimate the Ne number fraction from the the O number fraction using a scaling factor of $1/5.2$ as was determined for the \ion{H}{ii} region NGC~346 within the SMC \citep{2019Valerdi}.

\section{Results and Discussion}

\begin{figure*}[htp]
    \centering
    \includegraphics[width=\textwidth]{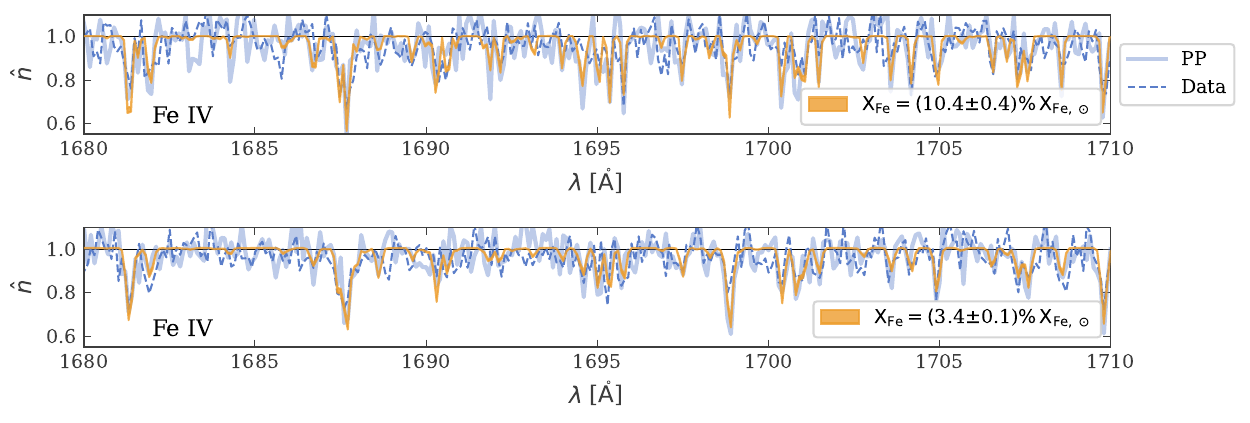}
    \caption{Comparison of the observed (dashed, blue line) and best-fit modeled (orange) \ion{Fe}{iv} forest of lines for fixed microturbulence values of $\xi=2\,\mathrm{km\,s^{-1}}$ for MBO2 (top) and $\xi=10\,\mathrm{km\,s^{-1}}$ for MBO3 (bottom). The thickness of the orange model line represents the $3\sigma$ credible interval of the best-fit model. A posterior predictive sample (PP) is shown in light blue. Full spectral fits are given in Appendix\,\ref{app:detailed_results}.}
    \label{fig:fits}
\end{figure*}

\subsection{The metal-poor O stars MBO2 and MBO3}\label{sec:MBO23}

In Fig.\,\ref{fig:fits}, we compare the best-fit models with the observed Fe forest for MBO2 and MBO3. 
The microturbulence $\xi$ is fixed to the values best reproducing the observed \ion{S}{v} line (cf.\ Fig.\,\ref{fig:vmic_comp}), i.e., $2~\mathrm{km\,s^{-1}}$ and $8~\mathrm{km\,s^{-1}}$ respectively, 
In addition, we show a posterior predictive sample, which is a set of simulated data points generated from the posterior distribution of model parameters. 
This sample allows us to assess the model's ability to reproduce the observed data.
For both stars, the simulated and observed data agree well, indicating that the model successfully reproduces the key features of the observations within the uncertainties. 
The corresponding full spectra and corner plots for MBO2 and MBO3 can be found in Appendix \ref{app:detailed_results}.

\begin{figure}[htp]
    \centering
    \includegraphics[width=\linewidth]{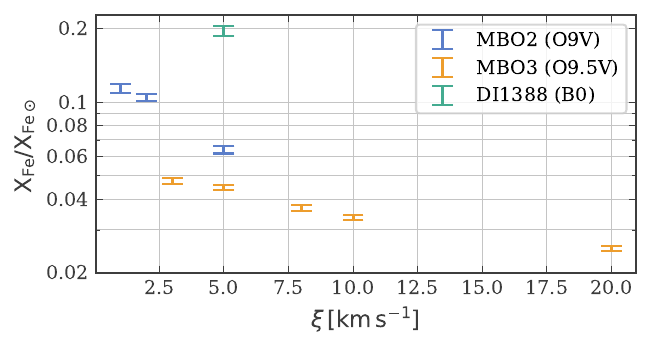}
    \caption{Fit results for the iron abundances of MBO2, MBO3 and DI1388 as a function of different microturbulence values $\xi$.}
    \label{fig:all_results}
\end{figure}

To study the degeneracy between $\xi$ and Fe abundance, we also performed the fit for other $\xi$-values. 
The detailed individual fitting results including the corner plots and corresponding spectra can be found in Appendix  \ref{app:detailed_results}.
A summary of the fitting results is depicted in Figure \ref{fig:all_results}. 
As expected, the value for the iron abundance decreases when we assume a larger microturbulence. 
The derived iron abundances differ by a factor of two depending on which microturbulence is assumed, even though they all fit the metal line profile by eye similarly well as shown in Fig. \ref{fig:vmic_comp}.

By combining the posterior distributions for the different fixed $\xi$ as described in Appendix \ref{app:microturb}, we obtain a final iron mass fraction of $1.41^{+0.10}_{-0.06}\times10^{-4}$ ($10.8^{+0.8}_{-0.5}~\%\,X_{\mathrm{Fe},\odot}$) for MBO2 and $4.79^{+0.10}_{-0.32}\times10^{-5}$ ($3.66^{+0.07}_{-0.25}~\%\,X_{\mathrm{Fe}, \odot}$) for MBO3. These values are lower than the adopted Fe abundance in the \citet{Ramachandran2021} models, but not different enough to affect the line predictions of other elements in the optical range.
Given the unaffected main stellar parameters and C, N, O, Mg, and Si abundances, we can thus combine our results with the abundances from \citet{Ramachandran2021} and our Ne assumption (cf.\ Sect.\,\ref{sec:bayes}) to obtain
a metallicity of $6.26\times10^{-4}$ ($\sim$$4.75 \% \, Z_\odot$) for MBO2 and $1.10\times10^{-3}$ ($\sim$$8.23 \% \, Z_\odot$) for MBO3. 
The derived flux shift values are within ${\text{5--7}\times10^{-15}\,\mathrm{erg\,cm^{-2}\,s^{-1}\, \AA^{-1}}}$ which corresponds to a shift of about 4--8\% of the measured UV flux and is in the same range as the measured flux uncertainties.

A model-independent, direct comparison of the Fe lines between S3 in Sextans A \citep{2024Telford} and MBO2 suggests a similar or lower Fe content for MBO2 (see Appendix \ref{sec:app-s3_comparison}).

\subsection{Comparison with the B star DI\,1388}\label{sec:DI1388}

To check for systematic differences with the metal-poor B star study by \citet{2008Dufton}, we analyzed DI~1388. We adopted the same stellar and broadening parameters from \citet{1994Hambly} as used by \citet{2008Dufton}. We calculated a grid of models for DI~1388 with varying Fe abundances and a fixed microturbulence of $\xi=5~\mathrm{km\,s^{-1}}$.
Instead of the plane-parallel TLUSTY code \citep{1988Hubeny}, we use the spherically-symmetric PoWR models and the same above-mentioned fitting procedure to derive the iron abundances.

In comparison to \citet{2008Dufton} ($X_{\mathrm{Fe}}=1.26 \% \, X_{\mathrm{Fe,\odot}}$), we determine a much larger iron abundance of $X_{\mathrm{Fe}}=(21 \pm 1) \% \, X_{\mathrm{Fe,\odot}}$ which is in the range of the mean SMC metallicity.
The comparison of the best-fit model with the data is shown in Fig. \ref{fig:DI1388} and the corresponding corner plot in Fig. \ref{fig:DI1388_corner}.
The difference in derived Fe abundance arises from a combination of factors, including differences in the i) underlined models ii) fitting methods, and iii) observational data. 
A major factor for the differences may be the different considered spectral wavelength ranges in \citet{2008Dufton} in comparison to this work.
In contrast to \citet{2008Dufton}, who used the range of $1890-1930\,\AA$ covering mainly the \ion{Fe}{iii} forest lines, we fit the models to the observed spectral range covering both the \ion{Fe}{iv} and \ion{Fe}{v} forest \citep[see, e.g.,][]{2020Hillier}.
At a temperature of $\sim$$32\,\mathrm{kK}$, \ion{Fe}{iv} and \ion{Fe}{v} are the dominant ions, which makes our chosen spectral range more suitable.
Additionally, the HST/GHRS spectra have a much lower signal-to-noise ratio ($\mathrm{S/N}=3$) in comparison to the STIS spectra ($\mathrm{S/N}=10-25$) used in this work.

\subsection{Abundance patterns and signs for $\alpha$ enhancement}

\begin{table*}
  \setlength{\tabcolsep}{10pt} 
  \renewcommand{\arraystretch}{1.2} 
    \centering
    \caption{Comparison of abundance measurements of hot massive stars and the ISM in the Magellanic Bridge. $[\mathrm{X/H}]$ denotes the logarithm of the number fractions of X and H relative to their solar values. $\Delta[\mathrm{X/H}]$ denotes the mean over all light metals (N, C, O, Si, Mg). The metallicity $Z$ is defined as the sum of all metal (N, C, O, Si, Mg, Ne, Fe) mass fractions. Spectral types of MBO2 and MBO3 are from \citet{Ramachandran2021}. }
    \begin{tabular}{ccccccc}
    \hline \hline\rule[-0.6ex]{0cm}{3.0ex}
        Star& Sp. Type & $\left[ \frac{\mathrm{Fe}}{\mathrm{H}}\right]$ & $\left[ \frac{\mathrm{O}}{\mathrm{H}}\right]$ & $\Delta \left[\frac{\mathrm{X}}{\mathrm{H}}\right]$ & $Z$ & Source \\ \hline
        MBO2        & O9V   & $-0.97$ & $-1.43$   &    $-1.35$   & $6.26\times10^{-4}$  &    This work \\
        MBO3        & O9.5V & $-1.44$ & $-0.98$   &    $-1.08$   & $1.10 \times10^{-3}$  &    This work\\
        DI 1388     & B0    & $-1.9$ &   $-0.5$   &    $<-0.5$ &   &    \citet{2008Dufton,1994Hambly} \\
                    &       &   $-0.68$   &   &            &   &    This work \\
        ISM         &       &  $-0.61$   &   $-0.96$     &   &   & \citet{2008Lehner} \\     
        \hline
    \end{tabular}
    \label{tab:comparison2}
\end{table*}
Our derived Fe abundances of O stars along with light metal abundances estimated from optical spectra \citep{Ramachandran2021} suggest that they exhibit distinct chemical abundances despite their close spatial proximity  (projected distance of $\sim$${15\,\mathrm{pc}}$).
A summary of different metallicity measures of the analyzed stars can be found in Table \ref{tab:comparison2}. 
Both MBO2 and MBO3 are significantly metal-deficient, by more than a factor of two, compared to the average ISM metallicity reported by \citet{2008Lehner}. In contrast, the more distant B star DI\,1388 (projected distance $\sim$$3.3\,$kpc), exhibits SMC-like Fe and overall elemental abundances.
An important result is that the iron abundances of MBO2 and MBO3 do not scale with the abundances of oxygen and other light elements.
MBO2, although having the lowest mean abundance of light metals of $\Delta \left[\mathrm{X/H}\right] = -1.35$, has a higher fraction of iron, $\left[\mathrm{Fe/H}\right] = -0.97$.
In contrast, MBO3 has a higher mean abundance of light metals of $\Delta \left[\mathrm{X/H}\right] = -0.98$ but a significantly  lower iron content of $\left[\mathrm{Fe/H}\right] = -1.44$.

\begin{figure}[htp]
    \centering
    \includegraphics[width=\linewidth]{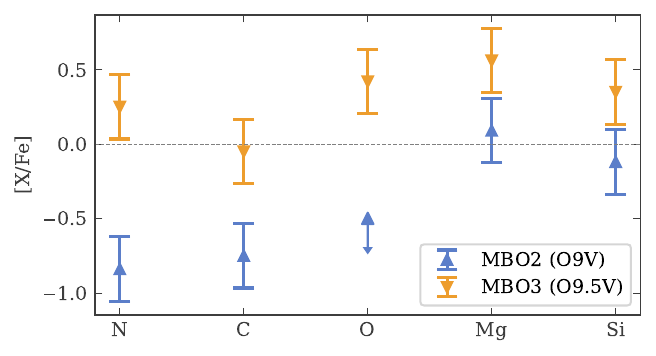}
    \caption{Depletion of metals relative to iron. For MBO2, only an upper limit could be determined for the oxygen abundance \citep{Ramachandran2021} which is highlighted by an arrow.}
    \label{fig:X-Fe}
\end{figure}

Figure \ref{fig:X-Fe} shows the X/Fe ratios (relative to solar values) for each element, including alpha elements like O, Mg, and Si, illustrating how elemental abundances scale with iron.
While $\alpha$-elements are mainly produced in massive stars and distributed to the ISM via their core-collapse supernovae, iron is primarily produced with a time delay through supernovae type Ia.
MBO3, which has the lowest iron content, shows signs for significant $\alpha$-enhancement. In MBO2, O and Si are depleted relative to Fe, whereas Mg exhibits a slight enhancement.

Thus, the inhomogeneous ISM of the Magellanic Bridge, suggested by \citet{Ramachandran2021} based on light element abundances, is further confirmed by our analysis. We find a large scatter in Fe abundances ($\sim 20-3\%\, Z_{\odot}$) and distinct [$\alpha$/Fe] ratios. These inhomogeneities may be linked to multiple gas accretion epochs, tidal interactions, or localized enrichment processes. To fully understand the complex chemical evolution of the Bridge, a more detailed and systematic chemical abundance analysis of a larger sample of young stars across the Bridge is essential (Sch\"osser et al., in prep). 

\section{Conclusions}

The low iron content of the stars and the inhomogeneities in their individual elemental abundances make the discovered O stars in the Magellanic Bridge unique counterparts to the high-redshift universe, where likely similar conditions existed.
Our results suggest that inhomogeneous mixing occurs even at the present epoch. At earlier times, mixing would have been even less efficient.
Thus, great caution is required when determining and interpreting abundances from integrated populations at high redshift. 
Already within the SMC and its outskirts, a variety of abundances is observed, most likely driven by interactions. 
At high redshift, these interactions were likely even more frequent, continuously changing the chemical content of the environment.
Moreover, our results highlight that even for young stars, there is no simple scaling between oxygen and iron and the inference from one element to the other can lead to significant over- or underestimations.
As iron is the main driver for winds and thus defines the main physics of the stars, these inaccuracies can lead to severe consequences for the inferred stellar evolution. This in turn leads to uncertainties in spectral energy distributions and ionizing radiation which affect chemical abundances derived from \ion{H}{ii} regions and contributions to cosmic reionization.

\begin{acknowledgements}
ECS acknowledges financial support by the Federal Ministry for Economic Affairs and Climate Action (BMWK) via the German Aerospace Center (Deutsches Zentrum f\"ur Luft- und Raumfahrt, DLR) grant 50 OR 2306 (PI: Ramachandran/Sander).
ECS further acknowledges support by the Federal Ministry of Education and Research (BMBF) and the Baden-Württemberg Ministry of Science as part of the Excellence Strategy of the German Federal and State Governments.
ECS, VR, AACS, MBP, and RRL acknowledge support by the German \emph{Deut\-sche For\-schungs\-ge\-mein\-schaft, DFG\/} in the form of an Emmy Noether Research Group -- Project-ID 445674056 (SA4064/1-1, PI Sander).
GGT acknowledges financial support by the Federal Ministry for Economic Affairs and Climate Action (BMWK) via the German Aerospace Center (Deutsches Zentrum f\"ur Luft- und Raumfahrt, DLR) grant 50 OR 2503 (PI: Sander).
GGT and JJ further acknowledge funding from the Deutsche Forschungsgemeinschaft (DFG, German Research Foundation) Project-ID 496854903 (SA4064/2-1, PI Sander). ECS, MBP, JJ, and RRL are members of the International Max Planck Research School for Astronomy and Cosmic Physics at the University of Heidelberg (IMPRS-HD).
\end{acknowledgements}

\bibliographystyle{aa} 
\bibliography{references} 

\appendix

\section{Data normalization and reduction}
\label{sec:app-basicred}

\begin{table*}
    \renewcommand{\arraystretch}{1.5} 
    \centering
    \caption{Details of the HST UV observations. The names and classifications of the stars are adopted from \citet{Ramachandran2021}.}
    \begin{tabular}{ccccccccc}
    \hline \hline
        Star    &  RA [deg] & DEC [deg] & Spectral type     &  Grating      & $\lambda$ $[\AA]$ & $R$   & Exposure time [s] & S/N\\ \hline
        MBO2    & 32.97933  & -74.21028 & O9V   & COS/G160\,M    & $1418-1789$   &   19000    & 1220 & 15-35 \\
        MBO3    & 33.02008  & -74.19933 & O9.5V & COS/G160\,M    & $1418-1789$   &   19000    & 1200  & 15-35 \\
        DI1388  &  44.29958 & -72.88194 & B0       &   STIS/E140\,M        &   $1140-1735$            &  45800     & 26160 & 10-25\\       \hline
    \end{tabular}
    \label{tab:coordinates}
\end{table*}

The UV spectra were obtained from Mikulski Archive for Space Telescopes (MAST) and are standard pipeline-processed data.
We normalize the observed spectra (listed in Table\,\ref{tab:coordinates}) by dividing them by a synthetic PoWR model continuum. As the shape of the spectral continuum does not change significantly for different iron metallicities between 1 and 15~\%, we use a model with a fixed iron group mass fraction of 5\% $G_\odot$ to normalize the UV spectra. 
Additionally, we applied a small flux shift to the synthetic model continuum. 
These shifts remained consistent with measured photometry.
For the Bayesian inference of the iron abundances (see Section \ref{sec:bayes}), we treated this flux shift as a free parameter in our fitting process, as it is difficult to distinguish between the uncertainty level and the absorption lines. Thus, instead of normalizing the spectrum once, we repeat the process for all iterations during the Bayesian sampling process. 

We then apply a radial velocity shift to the model spectra, $169\,\mathrm{km}\,\mathrm{s}^{-1}$ and $163\,\mathrm{km}\,\mathrm{s}^{-1}$ for MBO2 and MBO3 respectively. 
Furthermore, we account for galactic foreground and SMC reddening by applying the reddening law of \citet{1979Seaton} with $E_{\mathrm{B-V}} = 0.08$ to all model spectra. 
Additionally, we convolve the model spectra with a Gaussian with $\mathrm{FWHM} = 0.1$ to account for HST-COS's instrumental broadening, with a half-ellipse to account for rotation, and with a radial-tangential profile to account for macroturbulence.

\section{$\chi^2$ fit of the microturbulences}
\label{app:microturb}

\begin{figure}[htp]
    \centering
    \includegraphics[width=0.47\textwidth]{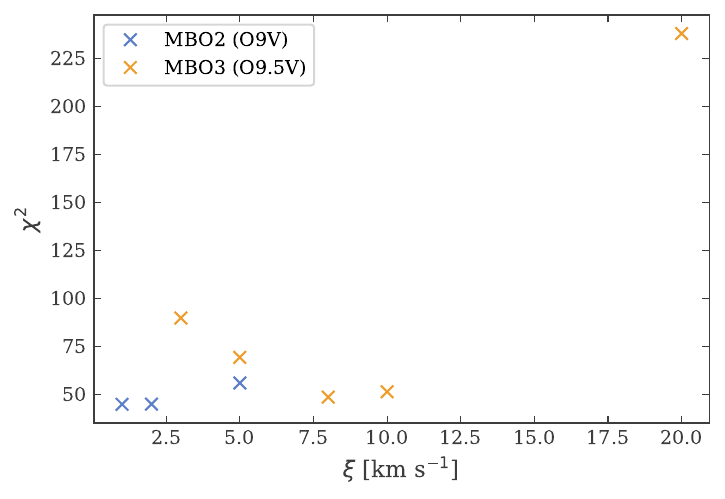}
    \caption{Computed $\chi^2$ values for the wavelength range corresponding to the shaded region in Fig. \ref{fig:vmic_comp} as a function of micro-turbulence for MBO2 and MBO3.}
    \label{fig:vmic_chisquared}
\end{figure}

We quantify the goodness of the fit for different microturbulences by computing 
\begin{align}
    \chi^2(\xi_i) = \sum_j \frac{(F_{\mathrm{obs, i}}- F_{\mathrm{sim, i}})^2}{\sigma_{\mathrm{i}}^2},
    \label{eq:chisquared}
\end{align}
which is the squared difference between the observed and model spectra, divided by the flux uncertainty $\sigma_{\mathrm{i}}^2$ at the $i$-th wavelength point. 
As we did not vary the sulfur abundance and we are only interested in the fit of the wings, we only compute $\chi^2$ for the wing region of the \ion{S}{v} line which we define by a velocity shift between $10$ and $60\,\mathrm{km\,s^{-1}}$ around the center of the line (see Fig. \ref{fig:vmic_comp}). 
The computed $\chi^2$ values for models with different $\xi$ for MBO2 and MBO3 are depicted in Fig. \ref{fig:vmic_chisquared}.

We derive the final iron abundances and associated uncertainties by considering the probability distribution of microturbulences. 
This is achieved by assigning normalized weights
\begin{align}
    w_i = \frac{\exp(- 0.5 \, \chi^2(\xi_i))}{\sum_i \exp(- 0.5 \, \chi^2(\xi_i))}
\end{align}
to each microturbulence where we use the $\chi^2$ value as derived from Equation \eqref{eq:chisquared}.
We combine the iron abundance posterior distributions as follows:
\begin{align}
    p(G) = \sum_i w_i \, p_i(G).
\end{align}

\section{Comparison with S3}
\label{sec:app-s3_comparison}
\begin{figure*}
    \centering
    \includegraphics[width=1\linewidth]{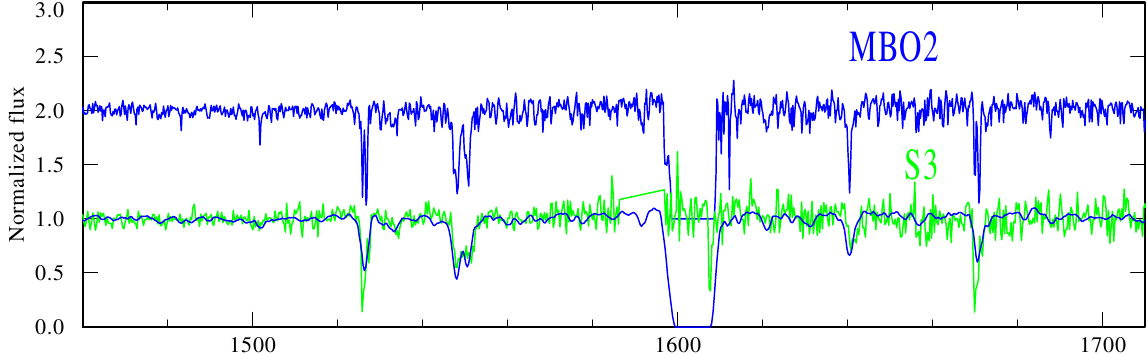}
    \caption{Comparison of the UV spectrum of MBO2 (blue) convolved with a rotational broadening profile ($\varv \sin i = 200$km/s, matching S3) with the observed UV spectrum of S3 (green). The original, unbroadened MBO2 spectrum (shifted upward for clarity) is also displayed.}
    \label{fig:s3convol}
\end{figure*}

As a model-independent comparison of the metal content, we compare the UV spectrum of MBO2 with the star S3 in Sextans A in Fig.\,\ref{fig:s3convol}, which have the same spectral type. To account for the significantly higher rotation of S3, we convolved the spectrum of MBO2 to match the observed $\varv \sin i = 200\,\mathrm{km\,s}^{-1}$ of S3. Despite the noise in the spectrum of S3, MBO2’s Fe lines appear similar to or weaker than those in S3, confirming the sub-SMC iron abundance of MBO2 and providing an independent consistency check of our new Fe abundance determination method.

\section{Atomic data}
\label{sec:app-atomicdata}

In Table\,\ref{tab:datom}, we list the number of levels and line transitions accounted for in the detailed non-LTE atmosphere calculations with PoWR.

\begin{table}[hp]
\caption{\label{tab:datom} Overview of the number of levels and lines per ion handled in non-LTE in the calculated atmosphere models.}
\begin{tabular}{lccc|lccr}
\hline\hline
Elem. & ion & levels & lines & Elem. & ion & levels & lines \\
\hline
 H    &     I     &  22  &   231  &   Ne                    &   VII   &   1   &     0    \\
 H    &    II     &   1  &     0  &   Al                    &    II   &  10   &    45    \\
 He   &     I     &  35  &   595  &   Al                    &   III   &  10   &    45    \\
 He   &    II     &  26  &   325  &   Al                    &    IV   &  10   &    45    \\
 He   &   III     &   1  &     0  &   Al                    &     V   &  10   &    45    \\
 N    &     I     &  10  &    45  &   Mg                    &     I   &   1   &     0    \\
 N    &    II     &  38  &   703  &   Mg                    &    II   &  12   &    66    \\
 N    &   III     &  36  &   630  &   Mg                    &   III   &  10   &    45    \\
 N    &    IV     &  38  &   703  &   Mg                    &    IV   &   1   &     0    \\
 N    &     V     &  20  &   190  &   Si                    &    II   &   1   &     0    \\
 N    &    VI     &  14  &    91  &   Si                    &   III   &  24   &   276    \\
 C    &     I     &  15  &   105  &   Si                    &    IV   &  23   &   253    \\
 C    &    II     &  32  &   496  &   Si                    &     V   &   1   &     0    \\
 C    &   III     &  40  &   780  &   P                     &    IV   &  12   &    66    \\
 C    &    IV     &  25  &   300  &   P                     &     V   &  11   &    55    \\
 C    &     V     &   5  &    10  &   P                     &    VI   &   1   &     0    \\
 C    &    VI     &  15  &   105  &   S                     &   III   &  23   &   253    \\
 O    &     I     &  13  &    78  &   S                     &    IV   &  11   &    55    \\
 O    &    II     &  37  &   666  &   S                     &     V   &  10   &    45    \\
 O    &   III     &  33  &   528  &   S                     &    VI   &   1   &     0    \\
 O    &    IV     &  25  &   300  &   G\tablefootmark{(S)}  &     I   &   1   &     0    \\                                                                                   
 O    &     V     &  36  &   630  &   G\tablefootmark{(S)}  &    II   &   3   &     2    \\                                                                                   
 O    &    VI     &  16  &   120  &   G\tablefootmark{(S)}  &   III   &  13   &    40    \\                                                     
 O    &   VII     &  15  &   105  &   G\tablefootmark{(S)}  &    IV   &  18   &    77    \\                                                     
 Ne   &     I     &   8  &    28  &   G\tablefootmark{(S)}  &     V   &  22   &   107    \\
 Ne   &    II     &   1  &     0  &   G\tablefootmark{(S)}  &    VI   &  29   &   194    \\
 Ne   &   III     &  18  &   153  &   G\tablefootmark{(S)}  &   VII   &  19   &    87    \\
 Ne   &    IV     &  35  &   595  &   G\tablefootmark{(S)}  &  VIII   &  14   &    49    \\
 Ne   &     V     &  54  &  1431  &   G\tablefootmark{(S)}  &    IX   &  15   &    56    \\
 Ne   &    VI     &  49  &  1176  &   G\tablefootmark{(S)}  &     X   &   1   &     0    \\
\hline                                                                                                                       
\end{tabular}
\tablefoot{%
    \tablefoottext{S}{Meta element containing the iron group elements Sc, Ti, V, Cr, Mn, Fe, Co, and Ni. The level and line numbers denote superlevels and superlines for this element, which are treated in non-LTE. Each superlevel contains a large set of explicit levels (of the same parity) within an energy band. Superlines describe the combined treatment of all transitions between two superlevels and contain a complex, pre-calculated cross section incorporating all explicit line transitions between the levels within the two superlevels \citep[see][for more details]{Grafener2002}.}
}
\end{table}

\section{Detailed fit results}\label{app:detailed_results}
\begin{figure*}[ph]
    \centering
    \includegraphics[width=0.5\textwidth]{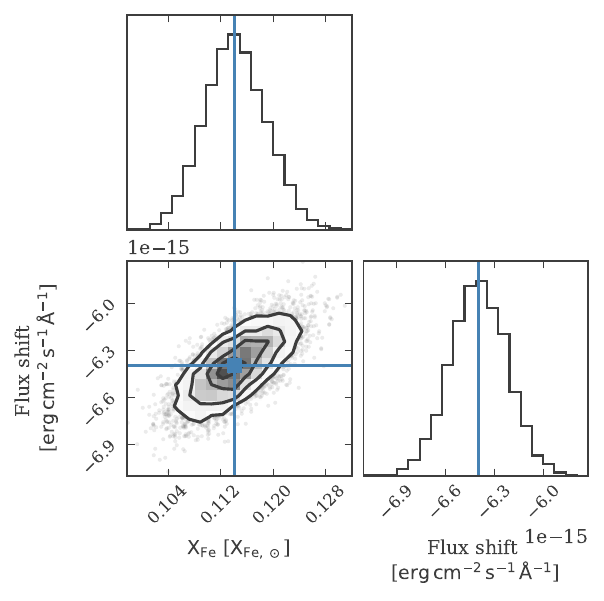}
    \caption{Corner plot for the fit of the iron mass fraction $\mathrm{X_{Fe}}$ and a flux shift in the spectral energy distribution for MBO2 and fixed microturbulence of ${\xi=1~\mathrm{km\,s^{-1}}}$.}
    \label{fig:O9V_vm1_corner}
\end{figure*}

\begin{figure*}[ph]
    \centering
    \includegraphics[width=0.8\textwidth]{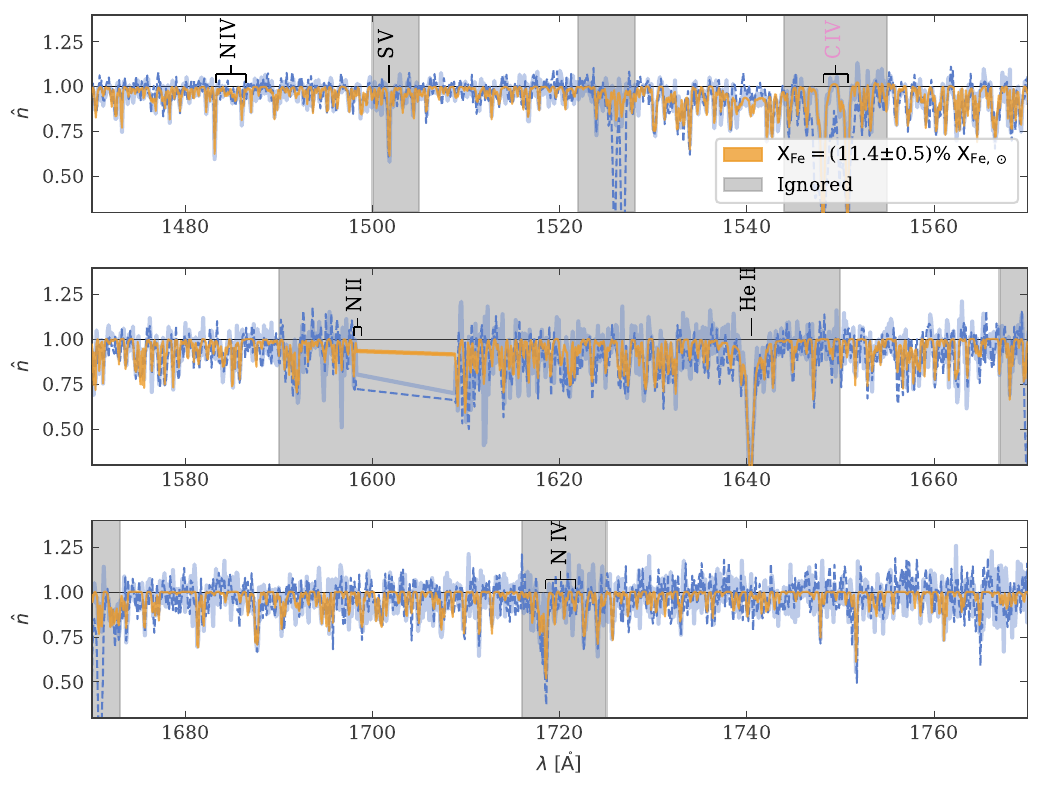}
    \caption{Comparison of the observed (dashed, blue line) and best-fit modelled spectrum (orange) for MBO2 and fixed microturbulence of ${\xi=1~\mathrm{km\,s^{-1}}}$. The thick light blue line depicts a posterior predictive sample. Regions which are ignored in the fit are highlighted in grey.}
    \label{fig:O9V_vm1}
\end{figure*}

\begin{figure*}[ph]
    \centering
    \includegraphics[width=0.5\textwidth]{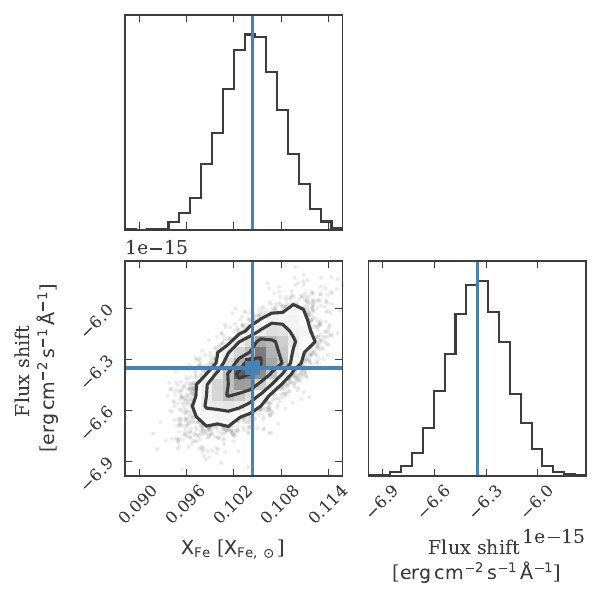}
    \caption{Corner plot for the fit of the iron mass fraction $\mathrm{X_{Fe}}$ and a flux shift in the spectral energy distribution for MBO2 and fixed microturbulence of ${\xi=2~\mathrm{km\,s^{-1}}}$.}
    \label{fig:O9V_vm2_corner}
\end{figure*}

\begin{figure*}[h]
    \centering
    \includegraphics[width=0.8\textwidth]{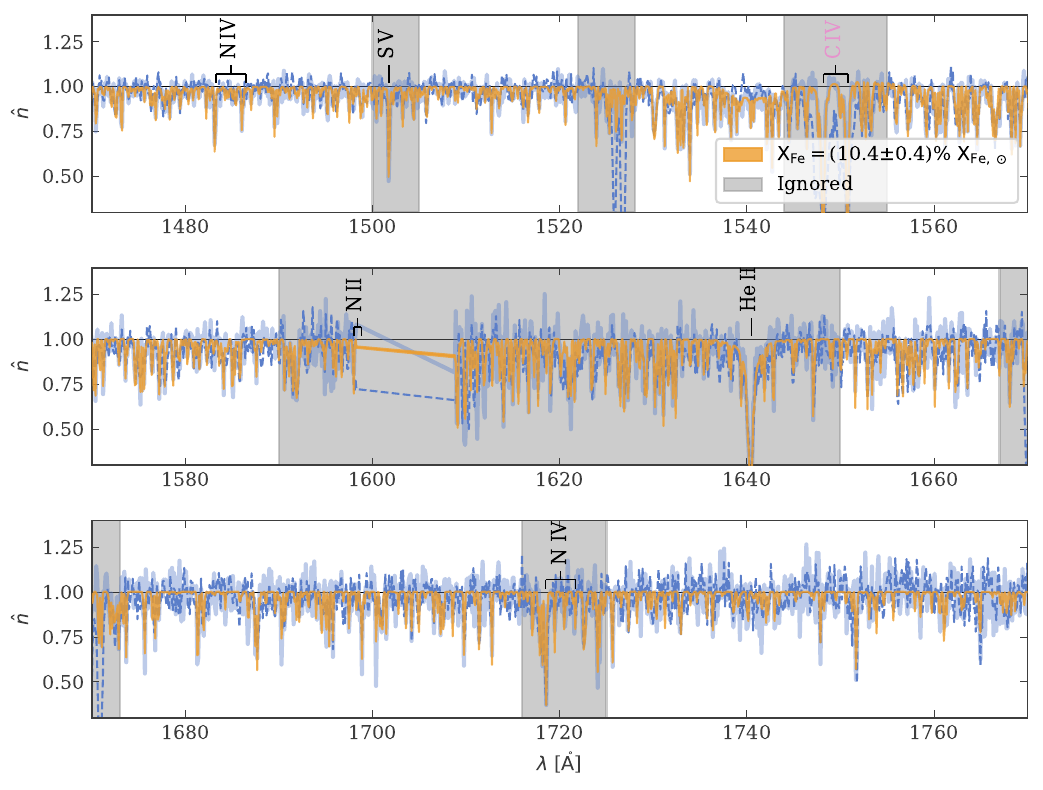}
    \caption{Comparison of the observed (dashed, blue line) and best-fit modelled spectrum (orange) for MBO2 and fixed microturbulence of ${\xi=2~\mathrm{km\,s^{-1}}}$. The thick light blue line depicts a posterior predictive sample. Regions which are ignored in the fit are highlighted in grey.}
    \label{fig:O9V_vm2}
\end{figure*}

\begin{figure*}[h]
    \centering
    \includegraphics[width=0.5\textwidth]{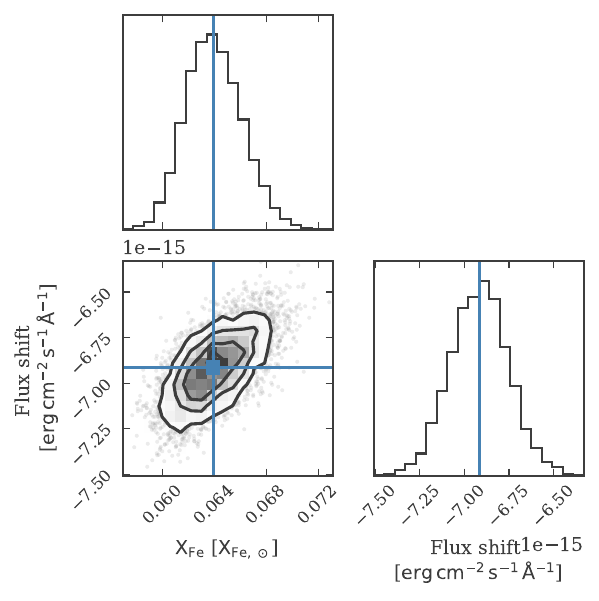}
    \caption{Corner plot for the fit of the iron mass fraction $\mathrm{X_{Fe}}$ and a flux shift in the spectral energy distribution for MBO2 and fixed microturbulence of $\xi=5~\mathrm{km\,s^{-1}}$.}
    \label{fig:O9V_vm5_corner}
\end{figure*}

\begin{figure*}[h]
    \centering
    \includegraphics[width=0.8\textwidth]{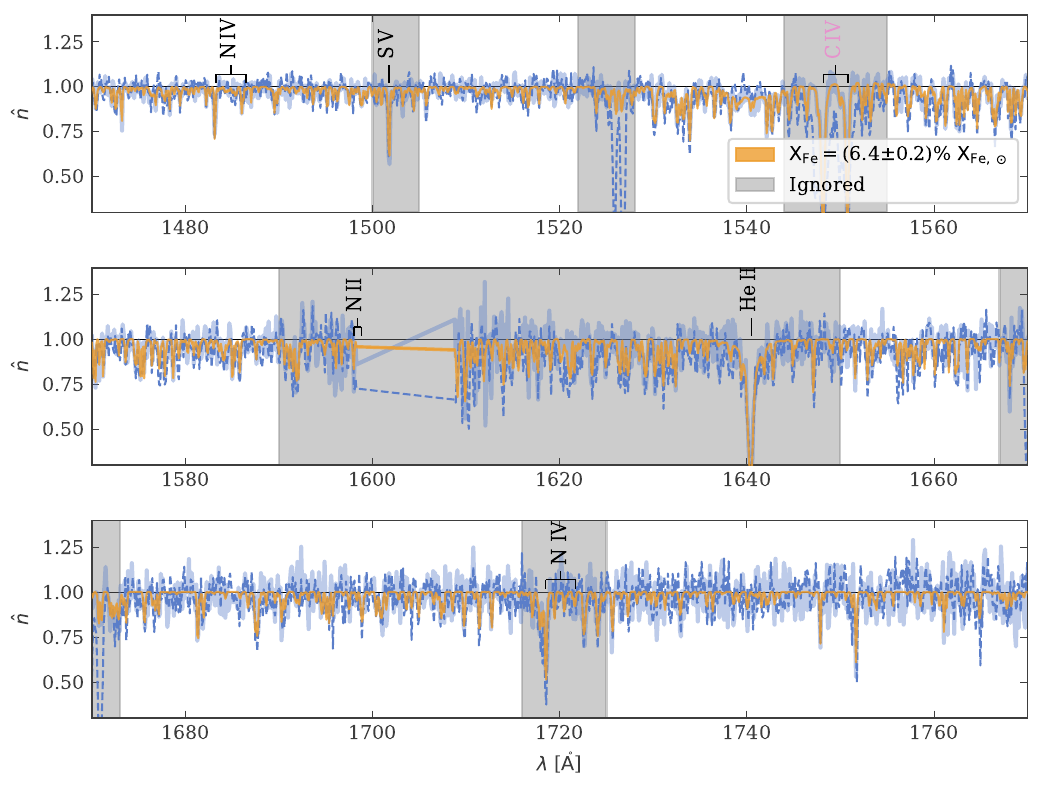}
    \caption{Comparison of the observed (dashed, blue line) and best-fit modelled spectrum (orange) for MBO2 and fixed microturbulence of ${\xi=5~\mathrm{km\,s^{-1}}}$. The thick light blue line depicts a posterior predictive sample. Regions which are ignored in the fit are highlighted in grey.}
    \label{fig:O9V_vm5}
\end{figure*}

\begin{figure*}[h]
    \centering
    \includegraphics[width=0.5\textwidth]{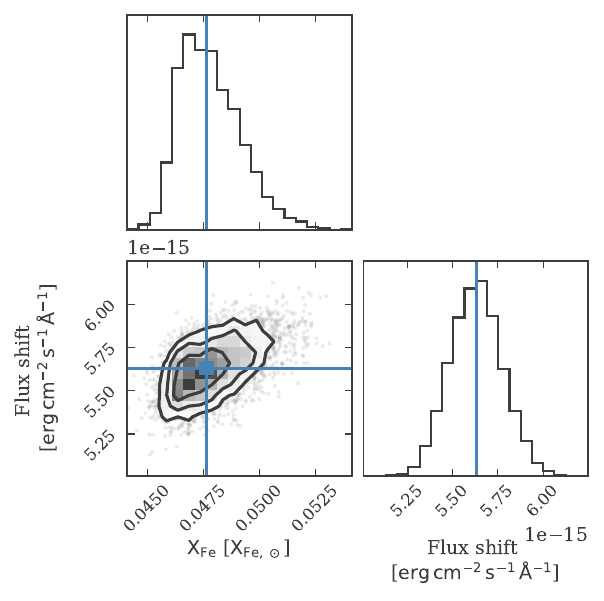}
    \caption{Corner plot for the fit of the iron mass fraction $\mathrm{X_{Fe}}$ and a flux shift in the spectral energy distribution for MBO3 and fixed microturbulence of $\xi=3~\mathrm{km\,s^{-1}}$.}
    \label{fig:O95V_vm3_corner}
\end{figure*}

\begin{figure*}[h]
    \centering
    \includegraphics[width=0.8\textwidth]{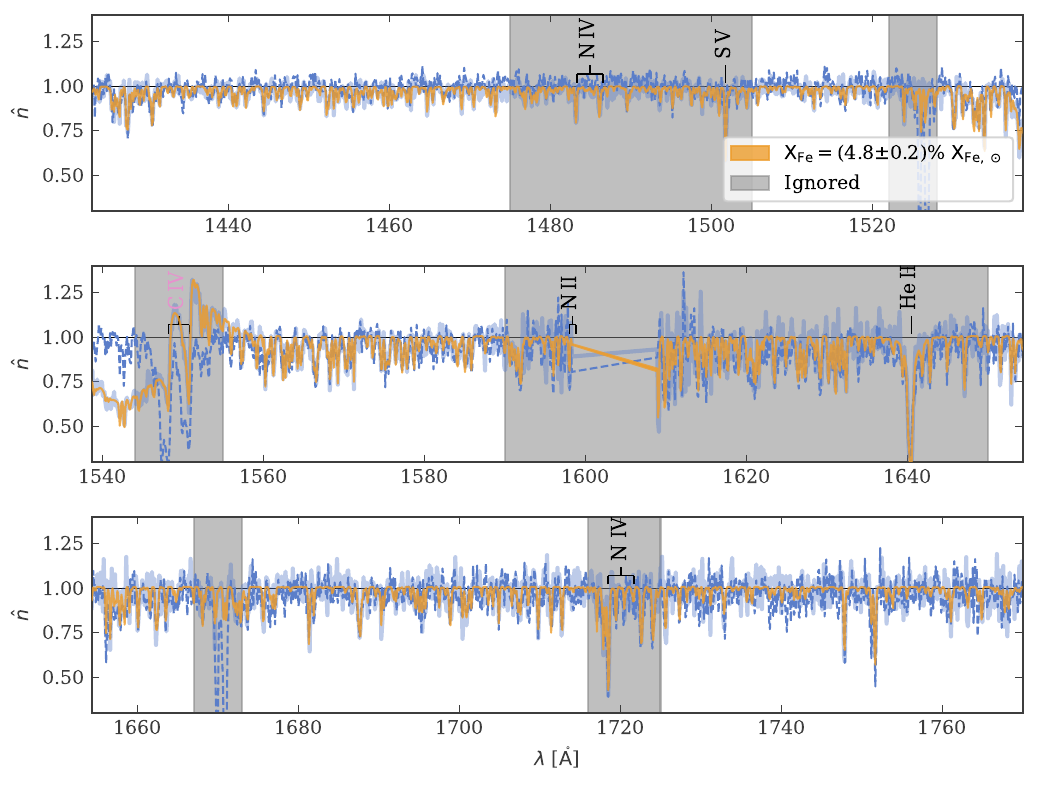}
    \caption{Comparison of the observed (dashed, blue line) and best-fit modelled spectrum (orange) for MBO3 and fixed microturbulence of ${\xi=3~\mathrm{km\,s^{-1}}}$. The thick light blue line depicts a posterior predictive sample. Regions which are ignored in the fit are highlighted in grey.}
    \label{fig:O95V_vm3}
\end{figure*}

\begin{figure*}[h]
    \centering
    \includegraphics[width=0.5\textwidth]{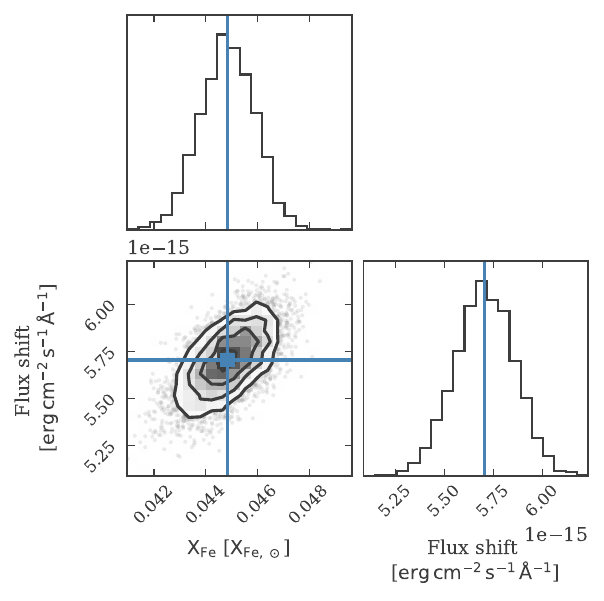}
    \caption{Corner plot for the fit of the iron mass fraction $\mathrm{X_{Fe}}$ and a flux shift in the spectral energy distribution for MBO3 and fixed microturbulence of $\xi=5~\mathrm{km\,s^{-1}}$.}
    \label{fig:O95V_vm5_corner}
\end{figure*}

\begin{figure*}[h]
    \centering
    \includegraphics[width=0.8\textwidth]{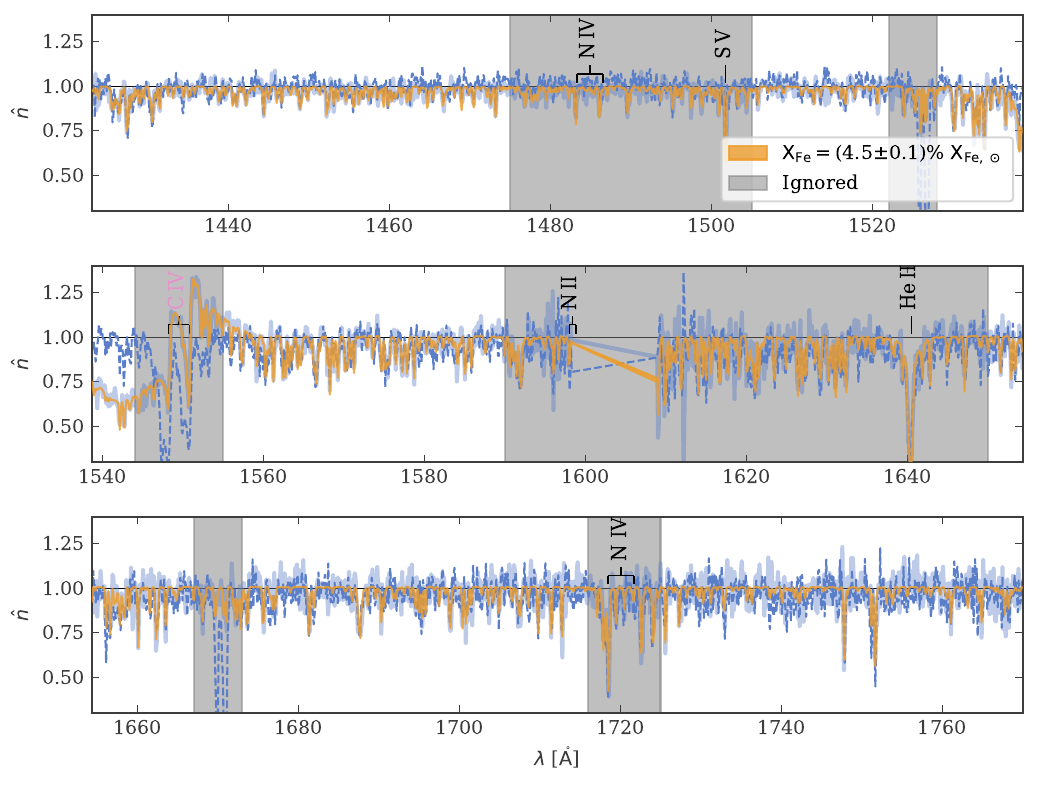}
    \caption{Comparison of the observed (dashed, blue line) and best-fit modelled spectrum (orange) for MBO3 and fixed microturbulence of $\xi=5~\mathrm{km\,s^{-1}}$. The thick light blue line depicts a posterior predictive sample. Regions which are ignored in the fit are highlighted in grey.}
    \label{fig:O95V_vm5}
\end{figure*}

\begin{figure*}[h]
    \centering
    \includegraphics[width=0.5\textwidth]{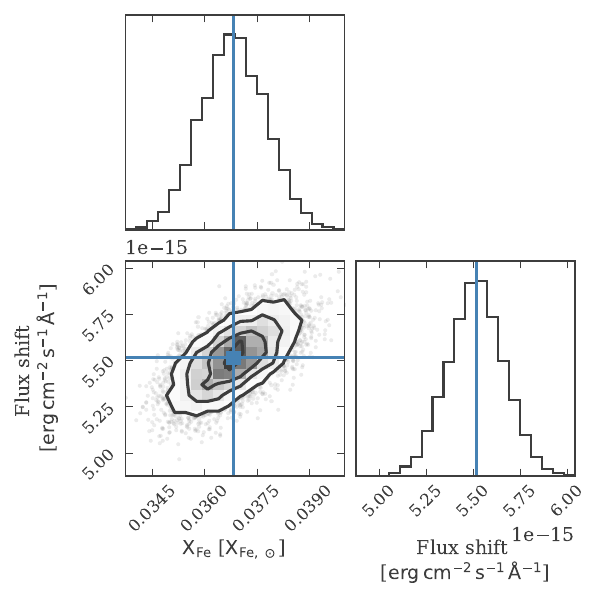}
    \caption{Corner plot for the fit of the iron mass fraction $\mathrm{X_{Fe}}$ and a flux shift in the spectral energy distribution for MBO3 and fixed microturbulence of $\xi=8~\mathrm{km\,s^{-1}}$.}
    \label{fig:O95V_vm8_corner}
\end{figure*}

\begin{figure*}[h]
    \centering
    \includegraphics[width=0.8\textwidth]{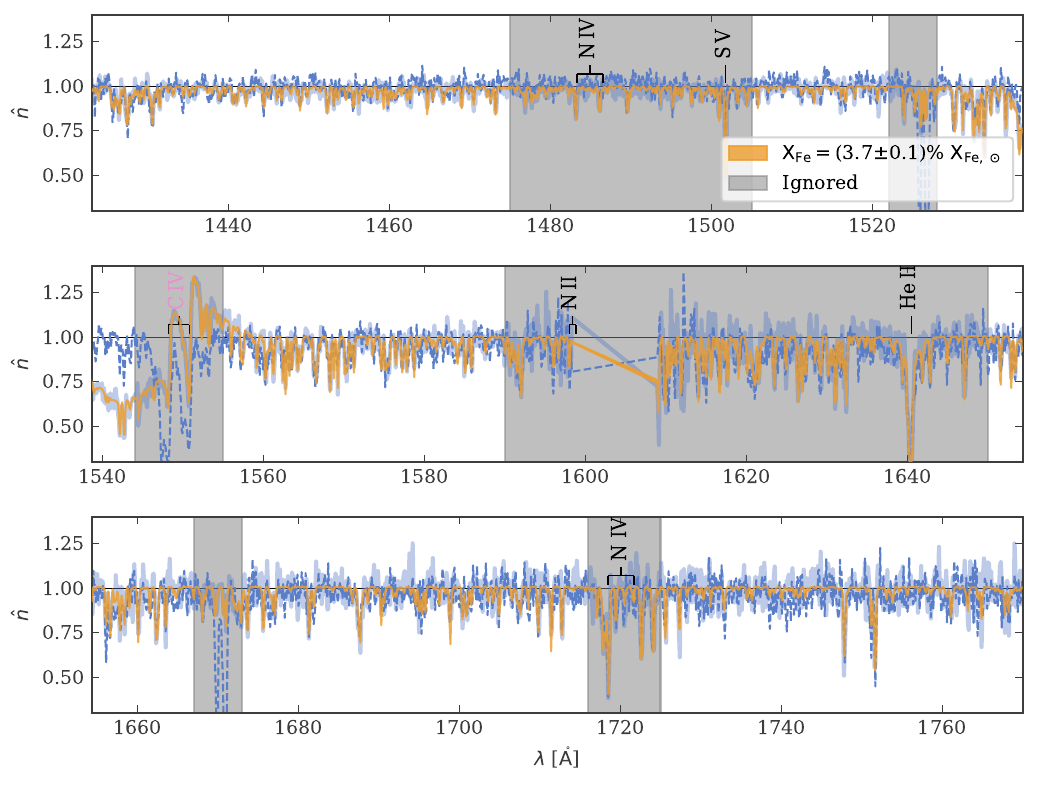}
    \caption{Comparison of the observed (dashed, blue line) and best-fit modelled spectrum (orange) for MBO3 and fixed microturbulence of $\xi=8~\mathrm{km\,s^{-1}}$. The thick light blue line depicts a posterior predictive sample. Regions which are ignored in the fit are highlighted in grey.}
    \label{fig:O95V_vm8}
\end{figure*}

\begin{figure*}[h]
    \centering
    \includegraphics[width=0.5\textwidth]{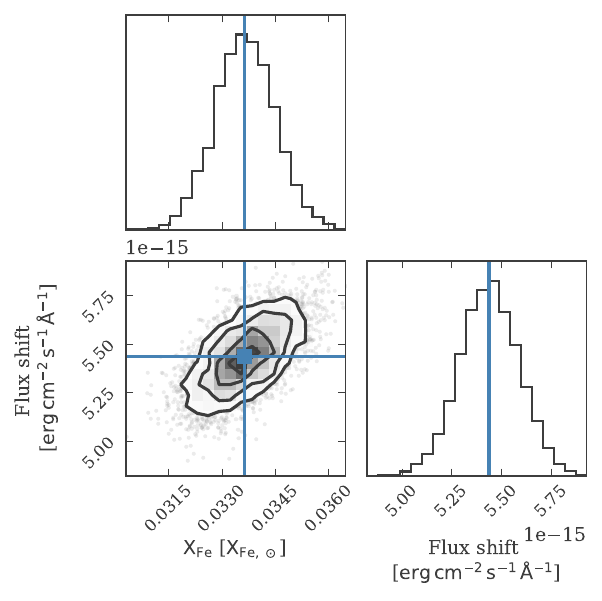}
    \caption{Corner plot for the fit of the iron mass fraction $\mathrm{X_{Fe}}$ and a flux shift in the spectral energy distribution for MBO3 and fixed microturbulence of $\xi=10~\mathrm{km\,s^{-1}}$.}
    \label{fig:O95V_vm10_corner}
\end{figure*}

\begin{figure*}[h]
    \centering
    \includegraphics[width=0.8\textwidth]{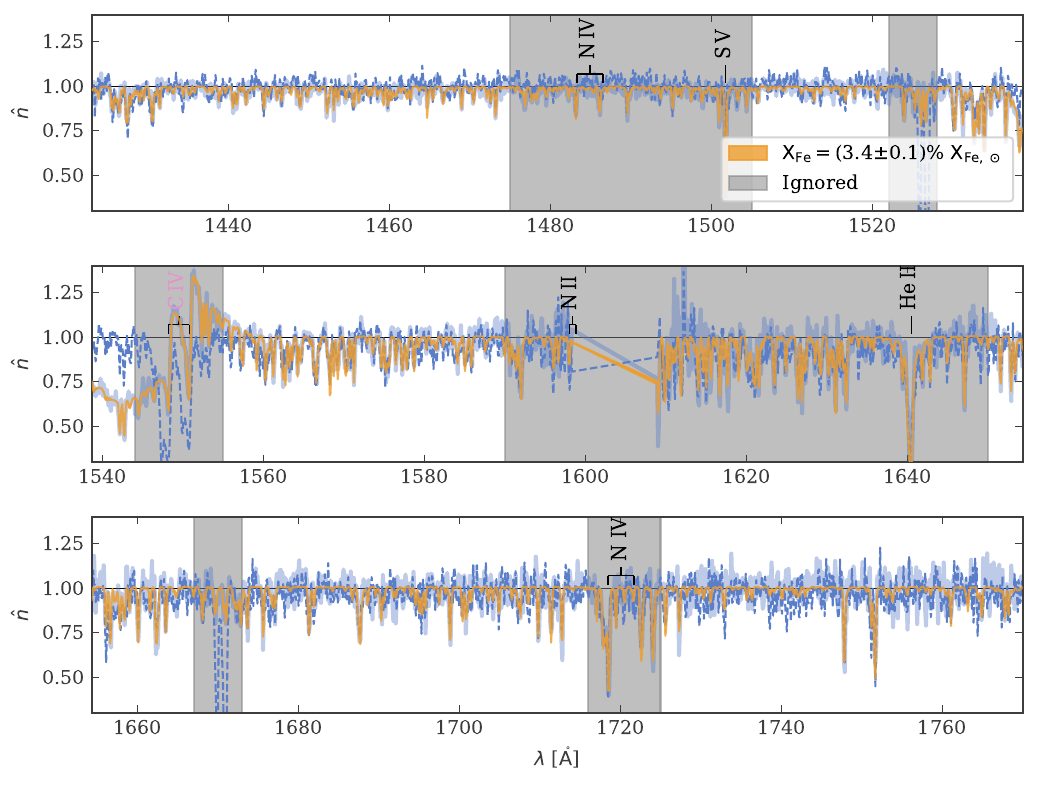}
    \caption{Comparison of the observed (dashed, blue line) and best-fit modelled spectrum (orange) for MBO3 and fixed microturbulence of $\xi=10~\mathrm{km\,s^{-1}}$. The thick light blue line depicts a posterior predictive sample. Regions which are ignored in the fit are highlighted in grey.}
    \label{fig:O95V_vm10}
\end{figure*}

\begin{figure*}[h]
    \centering
    \includegraphics[width=0.5\textwidth]{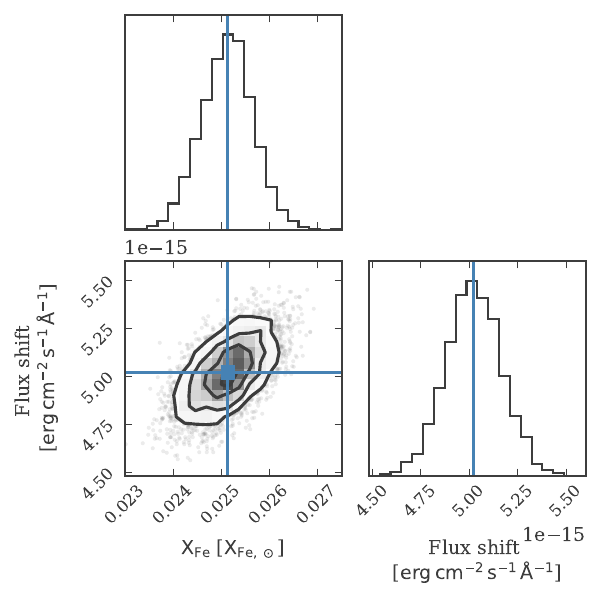}
    \caption{Corner plot for the fit of the iron mass fraction $\mathrm{X_{Fe}}$ and a flux shift in the spectral energy distribution for MBO3 and fixed microturbulence of $\xi=20~\mathrm{km\,s^{-1}}$.}
    \label{fig:O95V_vm20_corner}
\end{figure*}

\begin{figure*}[h]
    \centering
    \includegraphics[width=0.8\textwidth]{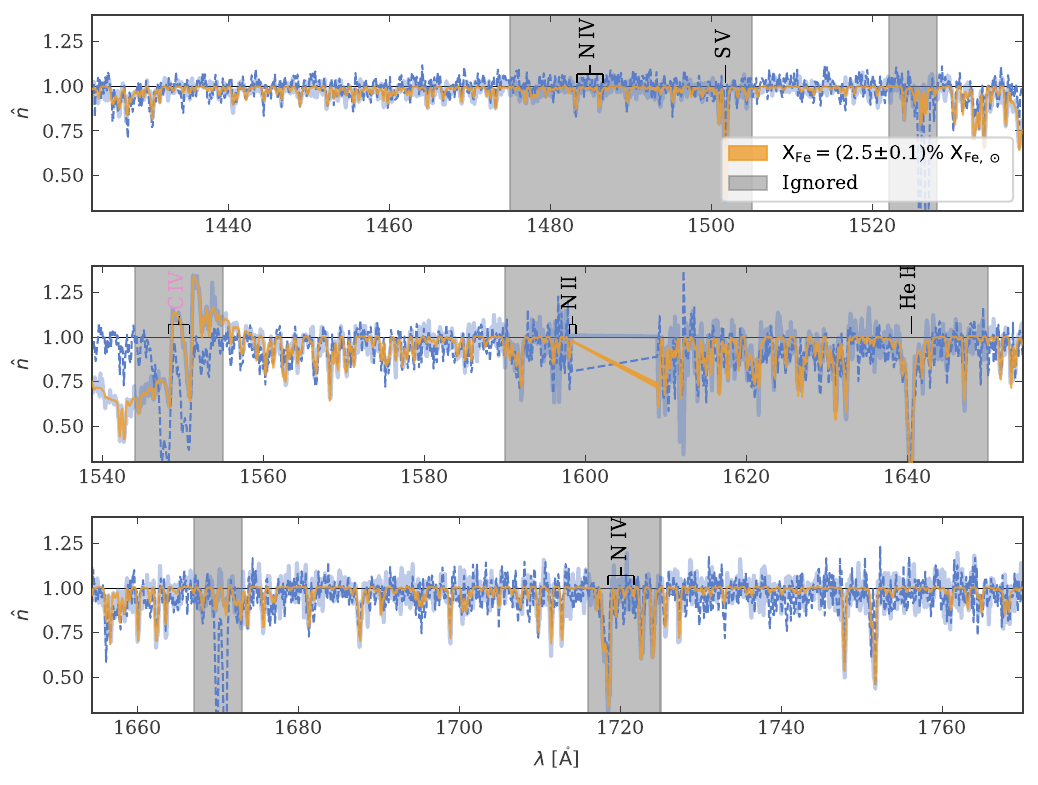}
    \caption{Comparison of the observed (dashed, blue line) and best-fit modelled spectrum (orange) for MBO3 and fixed microturbulence of $\xi=20~\mathrm{km\,s^{-1}}$. The thick light blue line depicts a posterior predictive sample. Regions which are ignored in the fit are highlighted in grey.}
    \label{fig:O95V_vm20}
\end{figure*}

\begin{figure*}[h]
    \centering
    \includegraphics[width=0.5\textwidth]{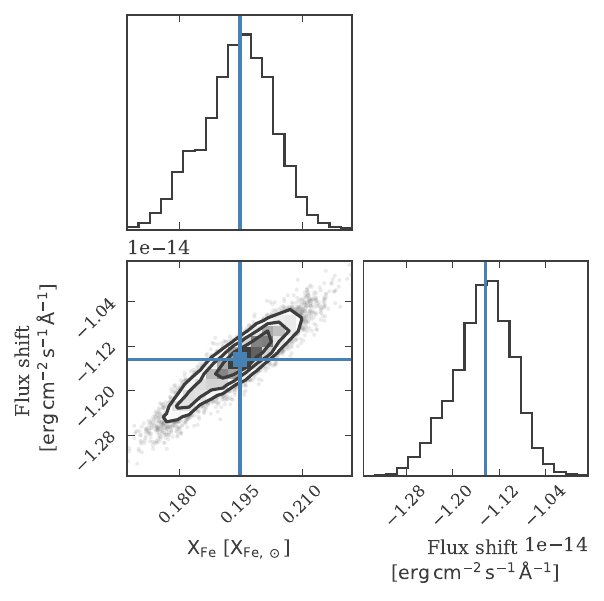}
    \caption{Corner plot for the fit of the iron mass fraction $\mathrm{X_{Fe}}$ and a flux shift in the spectral energy distribution for DI1388 and fixed microturbulence of $\xi=5~\mathrm{km\,s^{-1}}$.}
    \label{fig:DI1388_corner}
\end{figure*}

\begin{figure*}[h]
    \centering
    \includegraphics[width=0.8\textwidth]{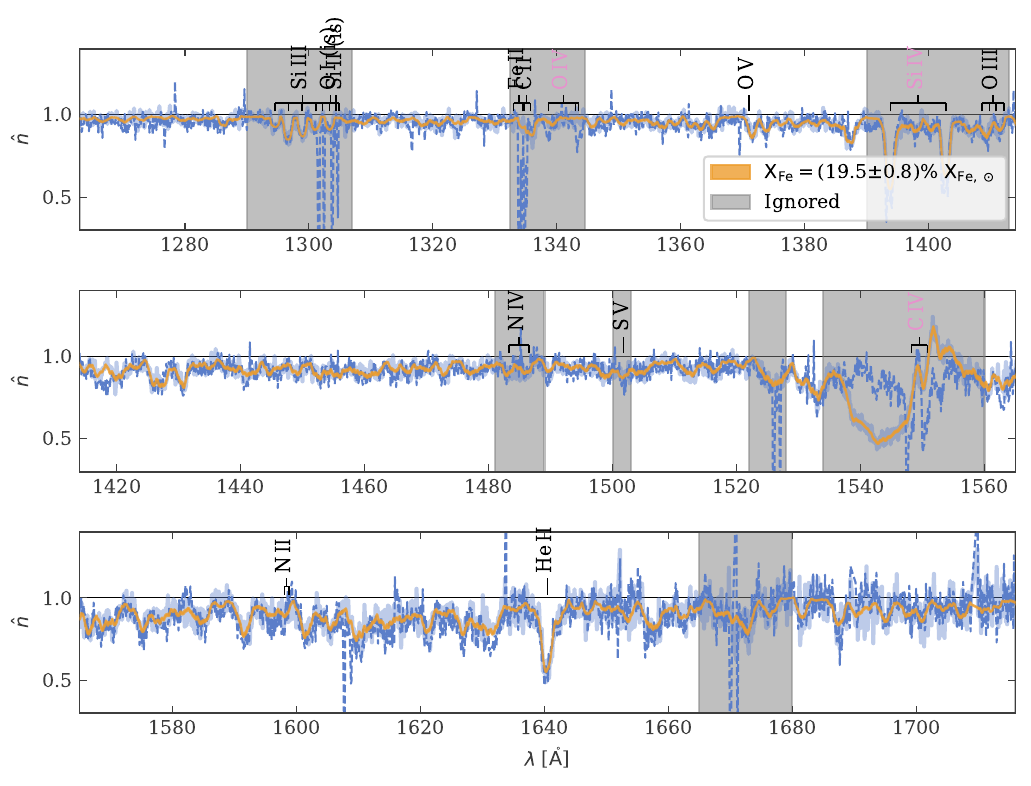}
    \caption{Comparison of the observed (dashed, blue line) and best-fit modelled spectrum (orange) for DI1388 and a fixed microturbulence of $\xi=5~\mathrm{km\,s^{-1}}$. The thick light blue line depicts a posterior predictive sample. Regions which are ignored in the fit are highlighted in grey.}
    \label{fig:DI1388}
\end{figure*}

In the following, we present the detailed fit results for MBO2, MBO3 and DI1388 including the corner plots and the best-fit model spectra over the full spectral range of HST/COS-G160M. 
The wavelength regimes that are ignored in the Bayesian analysis are highlighted by grey bands.
For MBO2, we show the individual fit results for the microturbulences $\xi=\text{1, 2 and 5}~\mathrm{km\,s^{-1}}$ (Figures \ref{fig:O9V_vm1_corner}--\ref{fig:O9V_vm5}), for MBO3 for $\xi=\text{3, 5, 8, 10 and 20}~\mathrm{km\,s^{-1}}$ (Figures \ref{fig:O95V_vm3_corner}--\ref{fig:O95V_vm20}), and for DI1388 for $\xi=\text{5}~\mathrm{km\,s^{-1}}$ (Figures \ref{fig:DI1388_corner}--\ref{fig:DI1388}).

\clearpage

\end{document}